# $h/2e$–Oscillations for Correlated Electron Pairs in Disordered Mesoscopic Rings

D. Weinmann, A. Müller-Groeling, J.-L. Pichard, and K. Frahm

*CEA, Service de Physique de l'Etat Condensé, Centre d'Etudes de Saclay,*
*91191 Gif-sur-Yvette Cedex, France*

The full spectrum of two interacting electrons in a disordered mesoscopic one–dimensional ring threaded by a magnetic flux is calculated numerically. For ring sizes far exceeding the one–particle localization length $L_1$ we find several $h/2e$–periodic states whose eigenfunctions exhibit a pairing effect. This represents the first direct observation of interaction–assisted coherent pair propagation, the pair being delocalized on the scale of the whole ring.

For more than three decades Anderson localization has been a subject of intensive research (for a review see [1]). Among the first results in this field was the fact that non-interacting electrons in one dimension (1d) are always localized and that their wave–functions decay exponentially on the scale of the localization length $L_1$. Very recently, Shepelyansky [2] and later Imry [3] considered two interacting electrons in a 1d random potential. Both authors suggest that in a regime of not too strong disorder correlated electron pair states may extend over a distance $L_2 \propto L_1^2$ much larger than $L_1$. This completely novel effect could very well have a fundamental impact on our understanding of the localization problem and other related areas. However, the methods employed in [2,3], though powerful and suggestive, are partly approximate and partly qualitative. Shepelyansky's original reasoning relied on certain assumptions of a statistical nature, allowing him to map the problem onto a random band matrix model. A numerical study of this model together with additional evidence from a variety of other models formed the basis for his claim. Imry, on the other hand, invoked the Thouless scaling block picture to arrive at precisely the same results as Shepelyansky. Moreover, Imry's approach seems to be very suitable to generalize the effect to higher dimensions.

The key quantity in [2,3] was the distribution of interaction matrix elements in the disorder–diagonal basis. This distribution defines the statistical properties of the band matrix model and provides the interblock coupling in the scaling approach. In a recent study [4] employing the transfer matrix technique we investigated numerically both finite and infinite systems starting from first principles. In this work, the principal effect was confirmed, but the two–particle localization length $L_2$ was found to scale with a smaller exponent, $L_2 \propto L_1^{1.65}$, in the parameter range investigated. This modification could be attributed to particular properties of the above-mentioned distribution which had not been taken into account in [2,3].

In the present paper, we demonstrate the existence of $h/2e$–periodic states in the spectrum of a disordered mesoscopic ring with two interacting electrons by direct diagonalization. Inspection of the two–electron eigenfunctions reveals that this period halving is indeed due to a pairing effect: Both electrons propagate coherently around the ring, staying within a distance of a few $L_1$ from each other. This is the first direct observation of interaction–assisted coherent pair propagation in disordered systems. The fact that the effect can be observed for the ring sizes considered (circumference $N = 50, 100, 150$) and the disorder chosen ($L_1 \approx 11$) supports our earlier investigations [4], where we found $L_2 \approx 25 > L_1$ for these system parameters and $U = 1$ (see (1) below). Here and in the remainder of the text we set the lattice spacing to unity. The flux sensitivities of the $h/2e$ states agree with what one can expect: an exponential reduction $\exp(-N/L_2)$ for $N > L_2$. Having calculated the full spectrum of the two–electron ring we also comment on the additional observation of interaction-enhanced $h/e$–oscillations. In the less localized case the enhancement simply originates from a screening effect, while sensitive $h/e$ states in the large rings benefit from a partial pairing effect. Our findings could also be relevant for the persistent current problem (see [5] and references therein).

To be specific, we consider a 1d ring which is threaded by the magnetic flux $\varphi = \Phi/\Phi_0$, where $\Phi_0 = h/e$ is the flux quantum. We employ a tight binding model with $N$ sites and random on-site energies $V_n$ ($n = 1, \ldots, N$), the latter being uniformly distributed in the interval $[-W, W]$. The electron–electron interaction is described by a local Hubbard term with strength parameter $U$. Thus the Hamiltonian reads

$$\mathcal{H} = \sum_{n=1}^{N} \sum_{\sigma} \left( e^{2\pi i\varphi/N} c^+_{n+1,\sigma} c_{n,\sigma} + e^{-2\pi i\varphi/N} c^+_{n,\sigma} c_{n+1,\sigma} \right) + \sum_{n=1}^{N} \sum_{\sigma} V_n c^+_{n,\sigma} c_{n,\sigma} + \sum_{n=1}^{N} U c^+_{n,\downarrow} c_{n,\downarrow} c^+_{n,\uparrow} c_{n,\uparrow}. \quad (1)$$

The operators $c^+_{n,\sigma}$ and $c_{n,\sigma}$ create and destroy an electron at site $n$ (we set $c^{(+)}_{N+1,\sigma} \equiv c^{(+)}_{1,\sigma}$) with spin $\sigma = \uparrow, \downarrow$, respectively.

We restrict our investigation to the case of two electrons with opposite spins. Then, the spatial part of the wave–function is symmetric, double–occupancy is al-lowed and the Hubbard interaction is relevant. We work in the basis of the $M = N(N+1)/2$ different wave–functions

$$| \psi_{n_1,n_2} \rangle | \sigma \rangle = \frac{1}{\sqrt{2}} \left( c^+_{n_1,\uparrow} c^+_{n_2,\downarrow} + c^+_{n_2,\uparrow} c^+_{n_1,\downarrow} \right) | 0 \rangle . \quad (2)$$

Since the Hamiltonian does not couple the spin components, we drop the antisymmetric spin part $| \sigma \rangle$ of the wave–function.

The resulting $M \times M$ Hamiltonian matrix is diagonalized by means of the Lanczos algorithm. The sensitivity of the two–electron energies $E_m$ $(m = 1, \ldots, M)$ on the flux as well as their periodicity is determined by calculating the full spectrum of $M$ eigenvalues of $\mathcal{H}$ for a few values of the magnetic flux between $\varphi = 0$ and $\varphi = 1/2$. For some levels with interesting flux dependence we also determine the corresponding eigenfunctions $| \Psi_m \rangle$. The shape of these eigenfunctions, i.e. the spatial distribution of the coefficients $\Psi_m(n_1, n_2) = \langle \psi_{n_1,n_2} | \Psi_m \rangle$, is very useful for the interpretation of the properties of the states and also helps to clarify the effects which lie at the origin of their behavior. Furthermore, we determine the "local" current

$$I_m(n_1, n_2) = \frac{e}{i} \left( \{ (\Psi^*_m(n_1+1, n_2) \Psi_m(n_1, n_2) \quad (3) \right.$$
$$\left. - \Psi^*_m(n_1, n_2) \Psi_m(n_1+1, n_2) \} + \{ n_1 \leftrightarrow n_2 \} \right) ,$$

corresponding to the center–of–mass motion of the two electrons. This quantity yields additional information about the way in which the two electrons propagate. The total current carried by the state $| \Psi_m \rangle$ is then given by $I^{tot}_m = \sum_{n_1=1}^{N} \sum_{n_2=n_1}^{N} I_m(n_1, n_2)$.

We briefly discuss the system parameters used in the sequel. Present–day computer technology restricts us to ring sizes of the order of $N \approx 10^2$. We have calculated full spectra for $N = 50, 100$, and $150$. The disorder parameter $W$ has to be chosen such that the ring is strongly localized in order to unambiguously identify coherent pair propagation. On the other hand, to see an effect at all we must have a sizable enhancement factor $L_2/L_1$ for the pair localization length $L_2$. We choose $W = 1.5$ and hence $L_1 = 25/W^2 \approx 11$ in the present work. For this value (and $U = 1$) we found $L_2 \approx 25$ in our transfer matrix study [4], i.e. an enhancement factor of more than two. For the Hubbard interaction we consider the values $U = -1, 0, 1$, as in [4]. We can estimate the amplitude $A$ of the $h/2e$ oscillations by the following considerations. For a ring of size $L_2$ roughly $L_1 L_2$ states will be coupled by the interaction, since a pair state has width $L_1$ and length $L_2$. The amplitude $A$ at scale $L_2$ is given by some fraction $c$ of the level spacing $\Delta_2(L_2)$ of the coupled states. The total bandwidth being $2(4+W)$, one gets $A \approx 2c(4+W)/L_1 L_2$. At scales $N > L_2$ this amplitude is suppressed by $\exp(-N/L_2)$. With $W = 1.5$ we therefore expect $A \approx c/200, c/1500, c/10000$ for the most sensitive states in rings with $N = 50, 100, 150$, respectively.

In Fig. 1 we compare the flux sensitivities of the energy levels for different ring sizes and different values of $U$. We have plotted the difference between the maximum and the minimum value in the interval $0 \leq \varphi \leq 1/2$ (i.e. the amplitude $A$) versus the energy value at zero flux. As the ring size increases the overall amplitudes decrease exponentially. More and more levels become localized and thus independent of the flux. Investigating the periodicity of the flux–sensitive levels, one finds for $U \neq 0$ that levels with strong $h/2e$ admixtures play a more and more prominent role in larger rings. These admixtures can be identified by the non–monotonous behavior of the eigenvalue in the flux interval $0 \leq \varphi \leq 1/2$ [6]. The amplitudes of the most sensitive $h/2e$ states are indicated in Fig. 1 by horizontal lines. For $N = 50$ the separation of the electrons forming a pair can be of the order of the ring circumference so that one expects $h/e$ rather than $h/2e$ oscillations. Indeed, very few $h/2e$ states exist for $N = 50$ and their amplitudes are dominated by those of the $h/e$ states by orders of magnitude. Most strikingly, this is no longer true for $N = 100$. There, $h/2e$–oscillations are less than an order of magnitude smaller than those of the largest $h/e$–periodic states. With an amplitude of about $10^{-4}$ the $h/2e$ states range between the comparatively few states with appreciable flux sensitivity, demonstrating the growing importance of the pair propagation process. The trend towards ever more pronounced $h/2e$–oscillations continues for $N = 150$. There, only very few of the $M = 11325$ levels are more sensitive to the flux than the $h/2e$–periodic ones. Assuming that some of the $h/e$ states for $N = 50$ are actually pair states we can compare our estimates for the amplitude $A$ with the numerical results. The exponential factor $\exp(-N/L_2)$ (and not $\exp(-N/L_1)$) is clearly confirmed by the data. Moreover, the absolute values are in good agreement for $c \approx 1/6$.

In Fig. 2 we show the flux–dependence for two selected $h/2e$ states, one for $N = 100$ and the other for $N = 150$. Density plots of the amplitudes $|\Psi_m(n_1, n_2)|^2$ of the corresponding wave–functions and the local currents $I_m(n_1, n_2)$ are presented in Fig. 3 and Fig. 4, respectively. They clearly exhibit the "cigar–shape" characteristic for coherent pair propagation: The wave–functions are concentrated around the diagonal $n_1 = n_2$ with a transverse extension given by a few $L_1$, but in the longitudinal (diagonal) direction they extend over the full ring. We recall that the circumference of the two rings is given by $9L_1$ and $14L_1$, respectively. The strong $h/2e$ admixture to the flux dependence of the eigenvalues together with the shape of the wave–function convincingly demonstrates that two electrons, staying within a distance $L_1$ of each other, propagate as a composite entity with charge $2e$. Moreover, the local current confirms and reinforces this interpretation. For instance, the branch along $n \approx 60$ in Fig. 4 is clearly identified as a dead end, showing that the electrons do not propagate around the ring independently. This is consistent with the almost perfect $h/2e$–periodic behavior of the corresponding energy level (see Fig. 2). We emphasize that the local current defined in (3) measures only the center–of–mass motion of the electrons. This quantity was chosen to simplify the presentation. We have checked that the local currents associated with the relative motion are similarly concentrated around the diagonal.

The behavior of the $h/e$–periodic states is also of interest. Comparing the flux–sensitivities in Fig. 1 for $U = 0$ and $U = 1$ one notices that while the average amplitude is slightly reduced for $U \neq 0$, a few very sensitive states emerge. This has been found for repulsive ($U = 1$) as well as for attractive ($U = -1$) interactions. This interaction–enhanced electron mobility is of a less subtle nature than the pair propagation effect. In Fig. 5 (left) we show a density plot of the wave–functions of particularly flux–sensitive $h/e$–periodic states for $U = 1$ and $N = 50$. This state exhibits a very pronounced cross–like structure, proving that the electrons propagate independently. The enhancement is simply due to screening: One of the electrons fills the deepest hole in the random potential landscape, occupying a state similar to the one–particle ground state $|\chi_0\rangle$. The second electron then moves in the smoothed effective potential $V_n^{eff} = V_n + |\chi_n|^2$, where $\chi_n$ is the component of $|\chi_0\rangle$ associated with site $n$. Therefore, this electron becomes delocalized and very sensitive to the magnetic flux. Similarly, for $U = -1$ the strongest positive fluctuation of the random potential is reduced by the attractive interaction between the delocalized electron and its localized partner, the latter occupying a one–electron state of high energy. This explains why highly flux–sensitive $h/e$ states of this type tend to occur in the upper (lower) half of the spectrum for $U = -1$ ($U = 1$). On the right hand side of Fig. 5, we show a particularly flux sensitive $h/e$ state for $N = 100$ and $U = 1$. Here, the comparatively high amplitude benefits from the pairing effect discussed above in the context of $h/2e$–periodic states. For this state, the pair is not extended over the whole ring, but splits up before each electron completes a cycle. The contribution of pair propagation in $h/e$ states for $N = 150$ is even more pronounced.

In conclusion, we have found $h/2e$–periodic states in the spectrum of a 1d disordered mesoscopic ring with two interacting electrons and circumference $N > L_1$. The corresponding wave–functions and local currents unambiguously demonstrate the pairing effect proposed by Shepelyansky. The amplitudes of these states agree well with simple estimates. The method, direct diagonalization, is free from any approximations whatsoever. The present investigation is in good agreement with our previous quantitative study of the pair localization length $L_2$ using the transfer matrix [4]. Furthermore we have observed an interaction–induced enhancement of $h/e$–oscillations due to the screening of the disorder potential in small rings, while pair propagation effects play an important role even for the $h/e$–oscillations in larger rings.

One might argue that the pair propagation effect is of limited relevance since we found only a few flux–sensitive states somewhere in the spectrum of the mesoscopic ring. This is, however, due to the relatively small system sizes to which we are restricted. For $N \approx L_2 \gg L_1 \gg 1$ the effects described in this paper should be much more dramatic. Furthermore, in a regime where the one–particle localization length $L_1$ at the edges of the one–particle band becomes much larger than the lattice spacing one can expect even the lowest–lying states of the ring to be affected. Also, both coherent pair propagation and the mechanism underlying the enhancement of $h/e$–oscillations might be relevant for the persistent current problem. In this case, one deals with a diffusive ring so that the maximum pair separation is larger than the ring size and one should not expect $h/2e$–oscillations. It is quite attractive to speculate that coherent pair propagation and the effective weakening of the disorder potential proposed in [5] are manifestations of the same physical effect at different scales. Much further work is needed to clarify these issues and to extend the results obtained up to now to higher dimensions and higher electron number.

We gratefully acknowledge fruitful discussions with Y. Imry and D. L. Shepelyansky. This work was supported by fellowships of the European HCM program (D. W., K. F.) and NATO/DAAD (A. M.-G.).

[1] B. Kramer and A. MacKinnon, *Rep. Prog. Phys.* **56**, 1469 (1993).
[2] D. L. Shepelyansky, *Phys. Rev. Lett.* **73**, 2067 (1994).
[3] Y. Imry, *preprint* (Rehovot, 1995).
[4] K. Frahm, A. Müller-Groeling, J.-L. Pichard and D. Weinmann, *preprint* (Saclay, 1995).
[5] A. Müller-Groeling, H. A. Weidenmüller and C. H. Lewenkopf, *Europhys. Lett.* **22**, 193 (1993); A. Müller-Groeling and H. A. Weidenmüller, *Phys. Rev.* **B 49**, 4752 (1994).
[6] For $N = 50$, some of the amplitudes are still of the order of the level spacing. The resulting avoided level crossings lead to the spurious impression of $h/2e$ admixtures in rather sensitive levels. Such levels have not been counted as $h/2e$–periodic.

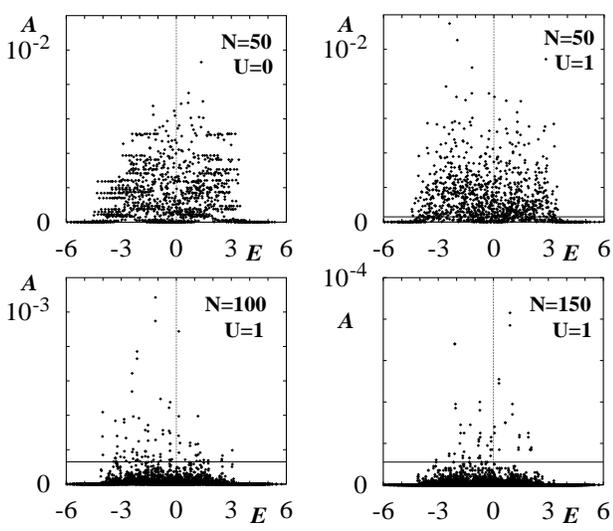

FIG. 1. The amplitudes $A$ of the oscillations of all levels as a function of the energy of the states. The upper pictures show the cases $U = 0$ (left) and $U = 1$ (right) for $N = 50$. The lower ones correspond to $N = 100$ (left) and $N = 150$ (right) at $U = 1$ (note the different scales for $A$). The total number of levels $M$ is 1275, 5050, and 11325 for the ring sizes 50, 100, and 150, respectively. The horizontal lines indicate the most sensitive $h/2e$-periodic levels.

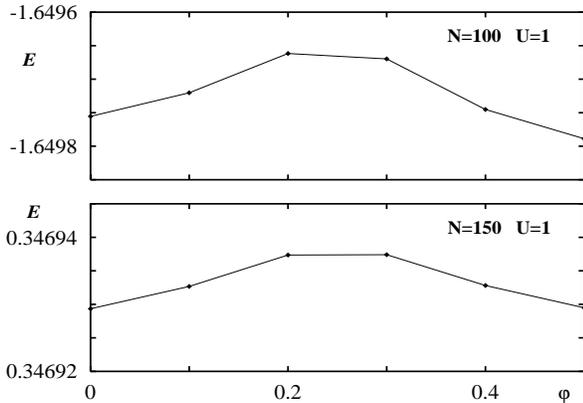

FIG. 2. The flux dependence of $h/2e$-periodic levels with high sensitivity for a ring of size $N = 100$ (top) and $N = 150$ (bottom) at $U = 1$. While the first one still has a small admixture of an $h/e$-periodic oscillation, the curve for the larger system exhibits almost perfect $h/2e$-periodicity.

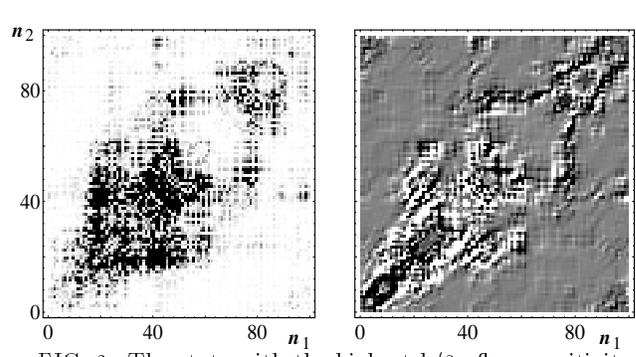

FIG. 3. The state with the highest $h/2e$ flux sensitivity at $N = 100$ and $U = 1$ corresponding to the level shown in Fig. 2 (top) for the flux value $\varphi = 0.15$. The absolute square of the wave-function is shown on the left-hand side. Dark regions correspond to high values. The associated local current is also shown (right). Here, dark regions correspond to positive values (propagation towards increasing $n_1$ and/or $n_2$), while bright regions indicate the opposite sign. In the grey areas, the current is close to zero.

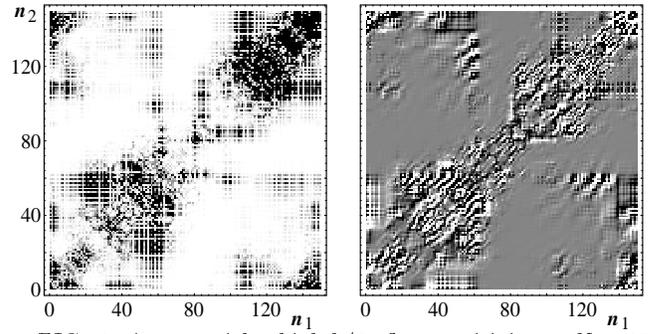

FIG. 4. A state with a high $h/2e$ flux sensitivity at $N = 150$ and $U = 1$ corresponding to the level shown in Fig. 2 (bottom). The absolute square of the wave-function (left) and the local current (right) are shown for $\varphi = 0.15$.

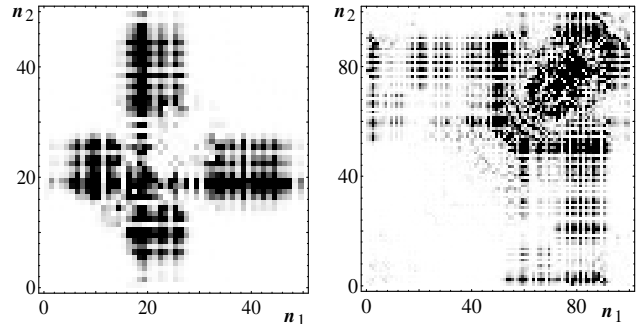

FIG. 5. The states with the highest flux sensitivities in rings containing $N = 50$ (left) and $N = 100$ (right) sites. They are found at $E \approx -2.411$ and $E \approx -1.139$, respectively. In the latter case a tendency towards a concentration near the diagonal indicates the formation of a pair, but the electrons split up before they complete a cycle.

4